\documentclass[12pt]{article}

\usepackage{amssymb}
\usepackage{cite}

\makeatletter
\renewcommand\thesubsection{\thesection.\@arabic\c@subsection}
\makeatother

\newcommand{\sect}[1]{\setcounter{equation}{0}\section{#1}}

\newcommand {\beq}{\begin{equation}}
\newcommand {\eeq}{\end{equation}}
\newcommand {\beqa}{\begin{eqnarray}}
\newcommand {\eeqa}{\end{eqnarray}}         %Equation version
\newcommand {\beqs}{\begin{eqnarray*}}
\newcommand {\eeqs}{\end{eqnarray*}}
\newcommand {\bds}{\begin{displaymath}}
\newcommand {\eds}{\end{displaymath}}
\newcommand {\n}{\nonumber\\}

\newcommand {\bebb}{\begin{thebibliography}}
\newcommand {\eebb}{\end{thebibliography}}      %Reference version
\newcommand {\bbit}{\bibitem}
\begin{document}

\newtheorem{Proposition}{Proposition}[section]
\newtheorem{Theorem}[Proposition]{Theorem}
\newtheorem{Definition}[Proposition]{Definition}
\newtheorem{Corollary}[Proposition]{Corollary}
\newtheorem{Lemma}[Proposition]{Lemma}
\newtheorem{Example}[Proposition]{Example}
\newtheorem{Remark}[Proposition]{Remark}

%\begin{titlepage}

\begin{flushright}
\end{flushright}

%\baselineskip =16pt

%\vskip.2in

\begin{center}
%\title
{\Large\bf Exact polynomial solutions of second order differential equations and their applications}

\vspace{0.5cm}

%\author{
{\large Yao-Zhong Zhang}
\vskip.1in

{\em School of Mathematics and Physics,
The University of Queensland, \\ Brisbane, Qld 4072, Australia}

%$b.$ {\em Institute of Modern Physics, Northwest University,
%Xi'an 710069, China}

\end{center}

\date{June 30, 2011}

%\maketitle

\vspace{1cm}

\begin{abstract}
\noindent We find all polynomials $Z(z)$ such that the differential equation
$$\left \{X(z)\frac{d^2}{dz^2}+Y(z)\frac{d}{dz}+Z(z)\right\}S(z)=0,$$
where $X(z), Y(z), Z(z)$ are polynomials of degree at most 4, 3, 2 respectively, has polynomial solutions
$S(z)=\prod_{i=1}^n(z-z_i)$ of degree $n$ with distinct roots $z_i$. We derive a set of $n$ algebraic equations
which determine these roots. We also find all polynomials $Z(z)$ which give polynomial solutions
to the differential equation when the coefficients of $X(z)$ and $Y(z)$ are algebraically dependent.
As applications to our general results, we obtain the exact (closed-form) solutions of the Schr\"odinger
type differential equations describing: 1) Two Coulombically repelling electrons on a sphere; 
2) Schr\"odinger equation from kink stability analysis of $\phi^6$-type field theory; 
3) Static perturbations for the non-extremal Reissner-Nordstr\"om solution; 
4) Planar Dirac electron in Coulomb and magnetic  fields;  and  5) $O(N)$ invariant decatic anharmonic oscillator.

\end{abstract}

\vskip.1in




%\vspace{0.5cm}

%\end{titlepage}

\setcounter{section}{0}
\setcounter{equation}{0}

\sect{Introduction and main results}

Consider the general 2nd order linear ordinary differential equation (ODE)
\beq
\left \{X(z)\frac{d^2}{dz^2}+Y(z)\frac{d}{dz}+Z(z)\right\}S(z)=0,\label{ODE}
\eeq
where $X(z), Y(z), Z(z)$ are polynomials of degree at most 4, 3, 2 respectively,
$$
X(z)=\sum_{k=0}^4a_kz^k,~~~~Y(z)=\sum_{k=0}^3b_kz^k,~~~~Z(z)=\sum_{k=0}^2c_kz^k.
$$
The ODE (\ref{ODE}) has as many as 12 parameters $a_k, b_k, c_k$, and contains,  as particular cases,
the Heun and generalized Heun equations as well as various confluent equations. As examples, here we list
five ODEs for each of which we provide a physical application in section \ref{applications}.
\begin{itemize}
\setlength{\itemsep}{0pt}
\setlength{\parskip}{0pt}
\setlength{\parsep}{0pt}
\item[(i)] $a_4=b_3=c_2=0$, i.e. deg$X(z)=3$, deg$Y(z)\leq 2$ and deg$Z(z)\leq 1$. 
If $X(z)$ has no multiple roots, we can write
%\beq
$X(z)=\prod_{s=1}^3(z-d_s),~~\frac{Y(z)}{X(z)}=\sum_{s=1}^3\frac{\alpha_s}{z-d_s}$
%\eeq
for suitable complex numbers $d_s, \alpha_s$. We then obtain the Heun equation,
\beq
\left\{\frac{d^2}{dz^2}+\sum_{s=1}^3\frac{\alpha_s}{z-d_s} \,\frac{d}{dz}
+\frac{Z(z)}{\prod_{s=1}^3(z-d_s)}\right\}S(z)=0. \label{Heun}
\eeq

\item[(ii)] deg$X(z)=4$, deg$Y(z)\leq 3$ and deg$Z(z)\leq 2$. If $X(z)$ has no multiple roots, we can write
%\beq
$X(z)=\prod_{s=1}^4(z-e_s),~~\frac{Y(z)}{X(z)}=\sum_{s=1}^4\frac{\mu_s}{z-e_s}$
%\eeq
for suitable c-numbers $e_s, \mu_s$. Then (\ref{ODE}) takes the form,
\beq
\left\{\frac{d^2}{dz^2}+\sum_{s=1}^4\frac{\mu_s}{z-e_s} \,\frac{d}{dz}
+\frac{Z(z)}{\prod_{s=1}^4(z-e_s)}\right\}S(z)=0. \label{gHeun1}
\eeq
This is the generalized Heun equation. 

\item[(iii)] deg$X(z)=3$ (i.e. $a_4=0$), deg$Y(z)\leq 3$ and deg$Z(z)\leq 2$. If $X(z)$ has no multiple roots, we can write
%\beq
$X(z)=\prod_{s=1}^3(z-f_s),~~\frac{Y(z)}{X(z)}=\sum_{s=1}^3\frac{\nu_s}{z-f_s}+\nu$
%\eeq
for suitable c-numbers $f_s, \nu_s, \nu$. Then (\ref{ODE}) has the form,
\beq
\left\{\frac{d^2}{dz^2}+\left(\sum_{s=1}^3\frac{\nu_s}{z-f_s} +\nu\right)\frac{d}{dz}
+\frac{Z(z)}{\prod_{s=1}^3(z-f_s)}\right\}S(z)=0. \label{gHeun2}
\eeq
This is the equation considered by Sch\"afke and Schmidt \cite{SS80}. 
 
\item[(iv)] deg$X(z)=2$ (i.e. $a_4=a_3=0$), deg$Y(z)\leq 3$ and deg$Z(z)\leq 2$. If $X(z)$ has no multiple roots, we can write
%\beq
$X(z)=(z-g_1)(z-g_2),~~\frac{Y(z)}{X(z)}=\frac{\sigma_1}{z-g_1}+\frac{\sigma_2}{z-g_2}+\sigma z+\kappa$
%\eeq
for suitable c-numbers $g_1, g_2, \sigma_1, \sigma_2, \sigma, \kappa$. Then (\ref{ODE}) is given by
\beq
\left\{\frac{d^2}{dz^2}+\left(\frac{\sigma_1}{z-g_1}+\frac{\sigma_2}{z-g_2}+\sigma z+\kappa\right)\frac{d}{dz}
+\frac{Z(z)}{(z-g_1)(z-g_2)}\right\}S(z)=0. \label{gHeun3}
\eeq

\item[(v)] deg$X(z)=1$ (i.e. $a_4=a_3=a_2=0$), deg$Y(z)\leq 3$ and deg$Z(z)\leq 2$. 
Write
%\beq
$\frac{Y(z)}{X(z)}=\frac{\eta}{z-h}+\lambda z^2+\gamma z+\delta$
%\eeq
for suitable c-numbers $h, \eta, \lambda, \gamma, \delta$. Then (\ref{ODE}) reads 
\beq
\left\{\frac{d^2}{dz^2}+\left(\frac{\eta}{z-h}+\lambda z^2+\gamma z+\delta\right)\frac{d}{dz}
+\frac{Z(z)}{(z-h)}\right\}S(z)=0. \label{gHeun4}
\eeq

%\item[(vi)] deg$X(z)=0$ (i.e. $a_4=a_3=a_2=a_1=0$), deg$Y(z)\leq 3$ and deg$Z(z)\leq 2$. We have
%\beq
%\left\{\frac{d^2}{dz^2}+\left(b_3z^3+b_2z^2+b_1 z+b_0\right)\frac{d}{dz} +Z(z)\right\}S(z)=0. \label{gHeun5}
%\eeq
\end{itemize} 
 
\vskip.2in
Recently there is a lot of research interest in finding polynomial solutions to second order 
differential equations of the form (\ref{ODE}) \cite{GKM09-10}-\cite{KUW96}.
%especially the Heun equation \cite{BS07,Shapiro-et-al-08-09}, 
%the generalized Heun equation \cite{SV03,MTV06} and certain special cases
%of (\ref{ODE}) in the context  of quasi-exact solvability (e.g. \cite{Turbiner96}-\cite{Dorey99}). 
ODEs with polynomial solutions are often called quasi-exactly solvable and
have wide-spread applications in physics, chemistry and engineering (see e.g. \cite{Ushveridze94} and references therein).  
One of the classical problems about the ODE (\ref{ODE}), suggested by E. Heine \cite{Heine1878}, T. Stieltjes
\cite{Stieltjes1885}  and G. Szego \cite{Szego39}, is \footnote{Authors in these references only considered the cases 
corresponding to (\ref{Heun}) and (\ref{gHeun1}), while the general ODE (\ref{ODE}) here also 
contains many other cases, e.g. differential equations (\ref{gHeun2})-(\ref{gHeun4}).}
\vskip.1in
\noindent {\bf Problem} {\em Given a pair of polynomials $X(z)$, $Y(z)$ and a positive integer $n$,
\begin{itemize}
\setlength{\itemsep}{0pt}
\setlength{\parskip}{0pt}
\setlength{\parsep}{0pt}
\item[(a)] find all polynomials $Z(z)$ such that the  ODE (\ref{ODE}) has a polynomial
solution $S(z)$ of degree $n$.
\item[(b)] find $S(z)$.
\end{itemize}
}

In this paper we solve this problem by means of the so-called Functional (or Analytic) Bethe Ansatz method \cite{Wiegmann94,Sasaki09,Yuan10,Yuan11}.
Precisely, we find the explicit values of the coefficients $c_2, c_1, c_0$ of
$Z(z)$ which give rise to degree $n$ polynomial solutions $S(z)$ of (\ref{ODE}) 
with roots $z_1, z_2,\cdots, z_n$ of multiplicity one, and we obtain a set of $n$ algebraic
equations (the so-called Bethe ansatz equations) which determine these roots..
%This has been used previously by us to find eigen-spectrum and polynomial solutions of certain types of higher-order ODEs.

For cases (i) and (ii) above with fixed real $d_s, e_s$ and positive $\alpha_s, \mu_s$ numbers in (\ref{Heun}) and (\ref{gHeun1}), respectively, 
there is a classical result, known as Heine-Stieltjes theorem \cite{Heine1878}. This theorem says \cite{Shapiro-et-al-08-09}
that if the coefficients of $X(z)$ and $Y(z)$ are algebraically independent, i.e. do not satisfy
any algebraic relations with integer coefficients, then for an arbitrary positive integer $n$ there are exactly 
$\left(\begin{array}{c}
n+{\rm deg}X(z)-2\\
n
\end{array}
\right) $ polynomials $Z(z)$ of degree exactly (deg$X(z)-2$) such that the ODE has a degree $n$ polynomial solution $S(z)$. 
However, even for these two cases, no results about the values of the coefficients $c_2, c_1, c_0$ of $Z(z)$ seem to be previously known. Furthermore, one may ask
\vskip.1in
\noindent{\bf Question} {\em If the coefficients of $X(z)$ and $Y(z)$ do satisfy some algebraic relations with
integer coefficients, i.e. are algebraically dependent, then how many polynomials $Z(z)$ are there which lead to 
degree $n$ polynomial solutions of the ODE (\ref{ODE})?  }  
\vskip.1in
To my knowledge, this question was not answered by Heine and Stieltjes in their theorem. 
In this paper we will also provide an answer to this question as a by-product of our general result (theorem \ref{main-results} below). 
 
We now state one of our main results of this paper.
\begin{Theorem}\label{main-results}
Given a pair of polynomials $X(z)$ and $Y(z)$, then the values of the coefficients $c_2, c_1, c_0$ of polynomial $Z(z)$ such that
the differential equation (\ref{ODE}) has degree $n$ polynomial solution 
\beq
S(z)=\prod_{i=1}^n(z-z_i)\label{polynomial-S}
\eeq
with distinct roots $z_1, z_2,\cdots,z_n$ are given by
\beqa
c_2&=& -n(n-1)a_4-nb_3,\label{c2-expression}\\
c_1&=&-\left[2(n-1)a_4+b_3\right]\sum_{i=1}^nz_i-n(n-1)a_3-nb_2,\label{c1-expression}\\
c_0&=&-\left[2(n-1)a_4+b_3\right]\sum_{i=1}^nz_i^2-2a_4\sum_{i<j}^nz_iz_j\n
   & &-\left[2(n-1)a_3+b_2\right]\sum_{i=1}^nz_i-n(n-1)a_2-nb_1,\label{c0-expression}
\eeqa
where the roots $z_1, z_2,\cdots, z_n$ satisfy the Bethe ansatz equations,
\beq
\sum_{j\neq i}^n\frac{2}{z_i-z_j}
+\frac{b_3z_i^3+b_2z_i^2+b_1z_i+b_0}{a_4 z_i^4+a_3z_i^3+a_2z_i^2+a_1z_i+a_0}=0,~~~~i=1,2,\cdots,n. \label{BAEs-general}
\eeq
The above equations (\ref{c2-expression})-(\ref{BAEs-general}) give all polynomials $Z(z)$ such that the ODE (\ref{ODE})
has polynomial solution (\ref{polynomial-S}).
\end{Theorem}

\begin{Remark}
T. Stieltjes gave in \cite{Stieltjes1885} a set of equations satisfied by the roots of $S(z)$ for
fixed real $\{d_s\}, \{e_s\}$ and positive $\{\alpha_s\}, \{\mu_s\}$ numbers in  (\ref{Heun}) and (\ref{gHeun1}), respectively. 
Our Bethe ansatz equations (\ref{BAEs-general}) reduce to those of Stieltjes under these conditions. 
However, even for these particular cases, the expressions for the coefficients of $Z(z)$, 
(\ref{c2-expression})-(\ref{c0-expression}), had not been obtained previously.   
Our theorem \ref{main-results} determines the explicit form of all polynomials $Z(z)$ which give polynomial solution to
the general ODE (\ref{ODE}). 
%not just the number of polynomials $Z(z)$ which allow polynomial solution to (\ref{ODE}).
{}From the explicit expressions of $c_2, c_1, c_0$ above, the number of $Z(z)$ is given by the number of the solution sets of 
the Bethe ansatz equations (\ref{BAEs-general}).   
\end{Remark}

{}From (\ref{c2-expression})-(\ref{BAEs-general}), we also have, as special cases of the theorem \ref{main-results},

\begin{Corollary} \label{by-products}
\begin{itemize}
\setlength{\itemsep}{0pt}
\setlength{\parskip}{0pt}
\setlength{\parsep}{0pt}
\item [(a)] If for an arbitrary integer $n$ the coefficients $a_4$ and $b_3$ in the ODE (\ref{ODE}) are algebraically dependent,
\beq
2(n-1)a_4+b_3=0,
\eeq
then there are  $n+1$ polynomials $Z(z)$ with coefficients $c_2=n(n-1)a_4, ~c_1=-n[(n-1)a_3+b_2]$ \cite{Turbiner96,Gonzarez93} and 
$$
c_0=-n\left[(n-1)a_2+2b_1)\right]-2a_4\sum_{i<j}^nz_iz_j-\left[2(n-1)a_3+b_2\right]\sum_{i=1}^nz_i,
$$
such that (\ref{ODE}) has degree $n$ polynomial solution $S(z)$ (\ref{polynomial-S}),
where the roots $z_i$ satisfy (\ref{BAEs-general}) with $b_3=-2(n-1)a_4$.  

\item[(b)] If $a_4=b_3=c_2=0$ and if for an arbitrary integer $n$ the coefficients $a_3$ and $b_2$ are algebraically
dependent,
\beq
2(n-1)a_3+b_2=0,
\eeq
then there is  $1$ polynomial $Z(z)=-n[(n-1)a_3+b_2]z-n(n-1)a_2-nb_1$ such that the corresponding 
ODE has degree $n$ polynomial solution $S(z)$ (\ref{polynomial-S}) with the roots
$z_i$ determined by (\ref{BAEs-general}) with $a_4=b_3=0,~ b_2=-2(n-1)a_3$.  
\end{itemize}

\end{Corollary}

%The results in part (b) of the Corollary is obvious from (\ref{c2-expression})-(\ref{BAEs-general}). The results 
%in part (a) will be proved in the next section by means of the $sl(2)$ algebraization of the corresponding ODE.

\vskip.1in
Other main results of this paper are the exact (closed-form) solutions of five physical systems 
(see section \ref{applications} below), 
providing interesting examples and applications of the differential equations (\ref{Heun})-(\ref{gHeun4}), respectively. 
The existence of such exact solutions demonstrates that these systems are quasi-exactly solvable.

In the appendix we present some explicit formulas corresponding to the differential equations (\ref{Heun})-(\ref{gHeun4}).

%\setcounter{section}{1}
%\setcounter{equation}{0}

%%%%%%%%%%%%%%%%%%%%%%%%%%%%%%%%%%%%%%%%%%%%%%%%%%%%%%%%%%%%%%%%%%%%%%%%%%%%%%%%%%%%%%%%%%%%%%%%%%%%%%%%%%%%%%%%%%%%%%%%%%%%%%%

\sect{The Proofs }

\noindent {\em Proof of Theorem \ref{main-results}.} We prove theorem \ref{main-results} by using the Functional Bethe ansatz method.
This method was used by us to obtain exact polynomial solutions of certain higher order ODEs arising from the nonlinear optical 
and spin-boson systems \cite{Yuan10,Yuan11}. Let
\beq
S(z)=\prod_{i=1}^n\left(z-z_i\right),\label{poly-solutions}
\eeq
be a degree $n$ polynomial  with undetermined, distinct roots $z_1, z_2,\cdots, z_n$. We will be
looking for the values of the coefficients $c_2, c_1, c_0$ of $Z(z)$ and the roots $z_i,~1\leq i\leq n$, such that (\ref{poly-solutions})
is a solution of (\ref{ODE}) with fixed coefficients of $X(z)$ and $Y(z)$. 
Substituting $Z(z)$ into the ODE and dividing on both sides by $Z(z)$ gives rise to
\beqa
-c_0&=&\left(a_4z^4+a_3z^3+a_2z^2+a_1z+a_0\right)\sum_{i=1}^n\frac{1}{z-z_i}\sum_{j\neq i}^n\frac{2}{z_i-z_j}\n
& &\left(b_3z^3+b_2z^2+b_1z+b_0\right)\sum_{i=1}^n\frac{1}{z-z_i}+c_2z^2+c_1z\nonumber
\eeqa
The left hand side is a constant and the right hand side is meromorphic function 
with simple poles at $z=z_i$ and singularity at $z=\infty$.
The residues of $-c_0$ at the simple poles $z=z_i$ are
\beqa
{\rm Res}(-c_0)_{z=z_i}&=&\left( a_4 z_i^4+a_3z_i^3+a_2z_i^2+a_1z_i+a_0\right)\sum_{j\neq i}^n\frac{2}{z_i-z_j}\n
& &+b_3z_i^3+b_2z_i^2+b_1z_i+b_0.\label{residues}
\eeqa
It can then be shown that
\beqa
-c_0-\sum_{i=1}^n\frac{{\rm Res}(-c_0)_{z=z_i}}{z-z_i}&=&
    \sum_{i=1}^n\left[a_4(z^3+z_i^3+z^2z_i+zz_i^2)+a_3(z^2+z_i^2+zz_i)\right.\n
& &\left. +a_2(z+z_i)+a_1\right]\sum_{j\neq i}^n\frac{2}{z_i-z_j}\n
& &+\sum_{i=1}^n\left[b_3(z^2+z_i^2+zz_i)+b_2(z+z_i)+b_1\right]+c_2z^2+c_1z\n
&=&\left[n(n-1)a_4+nb_3+c_2\right]z^2\n
& &+\left[\left(2(n-1)a_4+b_3\right)\sum_{i=1}^nz_i+n(n-1)a_3+nb_2+c_1\right]z\n
& &+\left(2(n-1)a_4+b_3\right)\sum_{i=1}^nz_i^2+2a_4\sum_{i<j}^nz_iz_j\n
& &+\left(2(n-1)a_3+b_2\right)\sum_{i=1}^nz_i+n(n-1)a_2+nb_1, \n
%\label{c0-residues}
\eeqa
where we have used identities,
\beqa
&&\sum_{i=1}^n\sum_{j\neq i}^n\frac{1}{z_i-z_j}=0,~~~~~
  \sum_{i=1}^n\sum_{j\neq i}^n\frac{z_i}{z_i-z_j}=\frac{1}{2}n(n-1),\n
&&\sum_{i=1}^n\sum_{j\neq i}^n\frac{z_i^2}{z_i-z_j}=(n-1)\sum_{i=1}^nz_i,~~~~~
  \sum_{i=1}^n\sum_{j\neq i}^n\frac{z_i^3}{z_i-z_j}=(n-1)\sum_{i=1}^nz_i^2+\sum_{i<j}^nz_iz_j.\nonumber
\eeqa
That is
\beqa
-c_0&=&\left[n(n-1)a_4+nb_3+c_2\right]z^2\n
& &+\left[\left(2(n-1)a_4+b_3\right)\sum_{i=1}^nz_i+n(n-1)a_3+nb_2+c_1\right]z\n
& &+\sum_{i=1}^n\frac{{\rm Res}(-c_0)_{z=z_i}}{z-z_i}\n
& &+\left(2(n-1)a_4+b_3\right)\sum_{i=1}^nz_i^2+2a_4\sum_{i<j}^nz_iz_j\n
& &+\left(2(n-1)a_3+b_2\right)\sum_{i=1}^nz_i+n(n-1)a_2+nb_1. \label{c0-residues}
\eeqa
%To eliminate the singularity of the right hand side of (\ref{c0-residues}) at $z=\infty$, we demand that 
The right hand side of (\ref{c0-residues}) is a constant if and only if the coefficients of  $z^2$ 
and $z$  as well as all the residues  at the simple poles are equal to zero, respectively. This gives $c_2, c_1$
in terms of the fixed coefficients of $X(z), Y(z)$ and the roots of $S(z)$, 
\beqa
&& n(n-1)a_4+nb_3+c_2=0,\n
&&\left(2(n-1)a_4+b_3\right)\sum_{i=1}^nz_i+n(n-1)a_3+nb_2+c_1=0,\nonumber
\eeqa
and the $n$ algebraic equations determining the roots $z_i$,
\beq
\sum_{j\neq i}^n\frac{2}{z_i-z_j}
+\frac{b_3z_i^3+b_2z_i^2+b_1z_i+b_0}{a_4 z_i^4+a_3z_i^3+a_2z_i^2+a_1z_i+a_0}=0,~~~~i=1,2,\cdots,n. \nonumber
\eeq
These equations are nothing but (\ref{c2-expression}), (\ref{c1-expression}) and  (\ref{BAEs-general}), respectively.
%Then the right hand side of (\ref{c0-residues}) becomes a constant. 
%On the other hand, the left hand side of (\ref{c0-residues}) if and only if 
%also becomes a constant if we demand
%that all the residues at the simple poles, (\ref{residues}), are zero. This gives the Bethe ansatz equations (\ref{BAEs-general}), 
% \beq
%\sum_{j\neq i}^n\frac{2}{z_i-z_j}
%+\frac{b_3z_i^3+b_2z_i^2+b_1z_i+b_0}{a_4 z_i^4+a_3z_i^3+a_2z_i^2+a_1z_i+a_0}=0,~~~~i=1,2,\cdots,n. \nonumber
%\eeq
It then follows from (\ref{c0-residues}) that
\beqa
-c_0& =&\left(2(n-1)a_4+b_3\right)\sum_{i=1}^nz_i^2+2a_4\sum_{i<j}^nz_iz_j\n
& &+\left(2(n-1)a_3+b_2\right)\sum_{i=1}^nz_i+n(n-1)a_2+nb_1, \nonumber
\eeqa
which is (\ref{c0-expression}). The proof is thus completed.

\hfill {$\Box$}

\vskip.2in
\noindent {\em Proof of Corollary \ref{by-products}}. Part (b) is obvious. We now prove part (a).
If the coefficients of $X(z)$ and $Y(z)$ are algebraically dependent,
$b_3=-2(n-1)a_4$, then the ODE (\ref{ODE}) has a hidden $sl(2)$ algebra symmetry if
$c_2=n(n-1)a_4, ~c_1=-n[(n-1)a_3+b_2]$.
To show this, we write the ODE (\ref{ODE}) with such coefficients $c_2, c_1$ as the Schr\"odinger form
\beq
H\,S(z)=-c_0 S(z). \label{H=-c0}
\eeq
Then it can be shown that $H$ is an element of the enveloping algebra of Lie-algebra $sl(2)$, 
\beqa
H&=&a_4J^+J^++a_3J^+J^0+a_2J^0J^0+a_1J^0J^-+a_0J^-J^-\n
 & &+\left[\frac{1}{2}(3n-2)a_3+b_2\right]J^+ +\left[(n-1)a_2+b_1\right]J^0\n
 & &+\left(\frac{n}{2}a_1+b_0\right)J^--\frac{n^2}{4}a_2+\frac{n}{2}\left[(n-1)a_2+b_1\right],\label{sl2-algebraization}
\eeqa
where
\beq
J^+=z^2\frac{d}{dz}-nz,~~~~J^0=z\frac{d}{dz}-\frac{n}{2},~~~~J^-=\frac{d}{dz}\label{sl2-differential}
\eeq
are differential operator realization of the $n+1$ dimensional representation of the $sl(2)$ algebra. 
$-c_0$ is the eigenvalue of $H$ with polynomial eigenfunction $S(z)=\prod_{i=1}^n(z-z_i)$, 
given by (from theorem \ref{main-results})
\beq
-c_0=n\left[(n-1)a_2+2b_1)\right]+2a_4\sum_{i<j}^nz_iz_j+\left[2(n-1)a_3+b_2\right]\sum_{i=1}^nz_i,\label{-c0-for-H}
\eeq
and the roots $z_i$ of the eigenfunction are determined by the Bethe ansatz equations,
\beq
\sum_{j\neq i}^n\frac{2}{z_i-z_j}
+\frac{-2(n-1)a_4z_i^3+b_2z_i^2+b_1z_i+b_0}{a_4 z_i^4+a_3z_i^3+a_2z_i^2+a_1z_i+a_0}=0,~~~~i=1,2,\cdots,n. \label{BAEs-for-H}
\eeq
It is well known that the solution space of second order differential operators with a $sl(2)$ algebraization is $n+1$ dimensional
\cite{Turbiner96,Gonzarez93}.  Applying this to the $H$ above, we conclude that the above Bethe ansatz equations
have $n+1$ sets of solutions and thus there $n+1$ eigenvalues $-c_0$, i.e. $n+1$ polynomials $Z(z)$. 

\hfill {$\Box$}
\vskip.1in
Let us remark that as far as we know the Schr\"odinger equation (\ref{H=-c0}) with $H$ given by (\ref{sl2-algebraization}) 
had not been solved previously (except for some very special cases), 
and thus (\ref{-c0-for-H}) and  (\ref{BAEs-for-H}) above give the first exact solution for the general ODE (\ref{H=-c0}).

\sect{Applications} \label{applications}

In this section we apply our general results obtained in section 1 to derive exact solutions of five physical systems, providing
examples and applications of the differential equations (\ref{Heun})-\ref{gHeun4}), respectively.

\subsection{Two Coulombically repelling electrons on a sphere} 

Consider a system of two electrons, interacting via a Coulomb potential, but constrained to remain on the surface of a
$D$-dimensional sphere of radius $R$ \cite{LG09}. The Hamiltonian of the system (in atomic units) is
\beq
H=-\frac{1}{2}\left(\nabla_1^2+\nabla_2^2\right)-\frac{1}{u},
\eeq
where $u=|{\bf r}_1-{\bf r}_2|$ is the inter-electronic distance. The Schr\"odinger wave function of the system can be
separated as a product of spin, angular and inter-electron wave functions. The inter-electron wave function $\Psi(u)$ 
satisfies the ODE \cite{LG09}
\beq
\left(\frac{u^2}{4R^2}-1\right)\frac{d^2\Psi}{du^2}+\left(\frac{\delta u}{4R^2}-\frac{1}{\gamma u}\right)\frac{d\Psi}{du}
   +\frac{\Psi}{u}=E\Psi,
\eeq
where $\delta$ and $\gamma$ are parameters related to the dimension $D$ of the sphere. Introduce dimensionless variable
$z=\frac{u}{2R}$. Then the above ODE can be written as
\beq
\left\{\frac{d^2}{dz^2}+\left(\frac{1/\gamma}{z}+\frac{\frac{1}{2}(\delta-1/\gamma)}{z+1}
  +\frac{\frac{1}{2}(\delta-1/\gamma)}{z-1}\right)\frac{d}{dz}+\frac{-4R^2Ez+2R}{z(z+1)(z-1)}\right\}\Psi=0.\label{heun-appl1}
\eeq
This ODE has the form  of the Heun equation (\ref{Heun}). It follows from our general results in previous sections that 
this equation has polynomial solutions of degree $n=1,2,\cdots,$
\beq
\Psi(z)=\prod_{i=1}^n(z-z_i),
\eeq
where $z_i$ are the roots of the above polynomial to be determined, provided that $E$ and $R$ take the values given by 
\beqa
E&=&\frac{n}{4R^2}(n+\delta-1),\label{energy-heun}\\
R&=&-\frac{1}{2}\left[2(n-1)+\delta\right]\sum_{i=1}^nz_i,\label{radius-heun} 
\eeqa
where $z_1,z_2,\cdots, z_n$ obey the Bethe ansatz equations
\beq
\sum_{j\neq i}^n\frac{2}{z_i-z_j}+\frac{1/\gamma}{z_i}+\frac{\frac{1}{2}(\delta-1/\gamma)}{z_i+1}
  +\frac{\frac{1}{2}(\delta-1/\gamma)}{z_i-1}=0,~~~~i=1,2,\cdots,n.\label{BAEs-2electrons}
\eeq
The energy eigenvalues $E$  agree with those obtained in \cite{LG09} by a different method. 
However, the above general, exact formula for the radius $R$ had not been given previously. 

For $n=1$, we have, from the Bethe ansatz equations, $z_1=\pm \frac{1}{\sqrt{\delta\gamma}}$. Then
$2R=-\delta z_1=\sqrt{\frac{\delta}{\gamma}}$ (by choosing the negative root $z_1=-\frac{1}{\sqrt{\delta\gamma}}$
so that the radius $R$ is non-negative), $E=\gamma$ and the wave function is 
$\Psi=z+\frac{1}{\sqrt{\delta\gamma}}=\frac{1}{\sqrt{\delta\gamma}}(1+\gamma u)$.

For $n=2$, we find 
\beqa
z_1&=&\frac{1}{2(\delta+2)}\left(-\sqrt{2(\delta+2)+\frac{4\delta+6}{\gamma}}\pm\sqrt{2(\delta+2)-\frac{2}{\gamma}}\right),\n
z_2&=&\frac{1}{2(\delta+2)}\left(-\sqrt{2(\delta+2)+\frac{4\delta+6}{\gamma}}\mp\sqrt{2(\delta+2)-\frac{2}{\gamma}}\right)
\eeqa
The other root satisfies $z_1=-z_2$ which leads to $R=0$ and therefore is discarded. The radius $R$ and the energy $E$ are
\beqa
R&=&-\frac{1}{2}(\delta+2)(z_1+z_2)=\frac{1}{2}\sqrt{2(\delta+2)+\frac{4\delta+6}{\gamma}},\n
E&=&\frac{\gamma(\delta+1)}{\gamma(\delta+2)+2\delta+3}
\eeqa
and the wave function is
\beq
\Psi=(z-z_1)(z-z_2)=\frac{1}{\gamma(\delta+2)}\left[1+\gamma u+\frac{\gamma^2(\delta+2)}{2\gamma(\delta+2)+4\delta+6}u^2\right].
\eeq

The ODE (\ref{heun-appl1}) has a hidden $sl(2)$ algebra symmetry. To see this, we rewrite (\ref{heun-appl1}) as
\beq
H\Psi=-2R\Psi,~~~~~H=z(z^2-1)\frac{d^2}{dz^2}+(\delta z^2-\frac{1}{\gamma})\frac{d}{dz}-4R^2Ez.
\eeq
If $E$ takes values given by (\ref{energy-heun}), i.e. $4R^2E=n(n-1+\delta)$, then $H$ is combination of $sl(2)$ algebra generators
(\ref{sl2-differential}),
\beq
H=J^+J^--J^0J^-+\left[\frac{1}{2}(3n-2)+\delta\right]J^+ -\left(\frac{1}{\gamma}+\frac{n}{2}\right)J^-.
\eeq
This provides an $sl(2)$ algebraization of the two electron system. This algebraization had not been realized previously.
The eigenvalues $-2R$ of $H$ with the polynomial eigenfunction $\Psi=\prod_{i=1}^n(z-z_i)$ are given (\ref{radius-heun})
and the roots $z_i$ are determined by the Bethe ansatz equations (\ref{BAEs-2electrons}).

\subsection{Schr\"odinger equation from the kink stability analysis of the $\phi^6$-type field theory}

Consider the $\phi^6$-type field theory in 1+1 dimensions characterized by the Lagrangian
\beq
{\cal L}=\frac{1}{2}\partial_\lambda\phi\partial^\lambda\phi-\frac{\mu^2}{8g^2(1+\epsilon^2)}
 \left(g^2\phi^2+\epsilon^2\right)\left(1-g^2\phi^2\right)^2,
\eeq
where $\epsilon$ is real dimensionless constant and $\mu$ has the dimension of mass.
As is well-known, the field equation of this theory has a kink solution. When one performs stability analysis
around the kink solution one arrives at the Schr\"odinger equation \cite{CL75,JKK89},
\beq
\left\{-\frac{d^2}{dx^2}+V(x)\right\}\psi(x)=E\psi(x),
\eeq
where $E\geq 0$ and the potential $V(x)$ is given by
\beq
V(x)=\mu^2\frac{8\sinh^4\frac{\mu x}{2}-\left(\frac{20}{\epsilon^2}-4\right)\sinh^2\frac{\mu x}{2}
    +2\left(\frac{1}{\epsilon^2}+1\right)\left(\frac{1}{\epsilon^2}-2\right)}
    {8\left(1+\frac{1}{\epsilon^2}+\sinh^2\frac{\mu x}{2}\right)^2}.
\eeq
We show that the Schr\"odinger equation can be transformed into the form of (\ref{gHeun1}). To this end,
we let
\beq
\psi=\left(1+\frac{1}{\epsilon^2}+\sinh^2\frac{\mu x}{2}\right)^{-\frac{3}{2}} y
\eeq
Then it can be shown that $y$ satisfies the ODE,
\beq
-y''+3\mu\frac{\sinh\frac{\mu x}{2}\cosh\frac{\mu x}{2}}{1+\frac{1}{\epsilon^2}+\sinh^2\frac{\mu x}{2}}y'
   +\frac{\frac{3}{2}\mu^2\left(1+\frac{1}{\epsilon^2}\right)}{1+\frac{1}{\epsilon^2}+\sinh^2\frac{\mu x}{2}}y
    =\left(E+\frac{5}{4}\mu^2\right)y.
\eeq
Make a change of variable,
\beq
z=\cosh\frac{\mu x}{2}
\eeq
Then the above equation becomes
\beqa
&&\left[z^4+\left(\frac{1}{\epsilon^2}-1\right)z^2-\frac{1}{\epsilon^2}\right]y''
  -\left[5z^3-\left(\frac{1}{\epsilon^2}+6\right)z\right]y'\n
&&~~~~~~~~~~~~~~~~~~~~~~
  +\left[\left(\frac{4E}{\mu^2}+5\right)z^2+\frac{4E}{\epsilon^2\mu^2}-\frac{1}{\epsilon^2}-6\right]y=0.\label{ODE-example2}
\eeqa
This ODE can be easily cast into the form of the generalized Heun equation (\ref{gHeun1}), noting that $X(z)=
z^4+\left(\frac{1}{\epsilon^2}-1\right)z^2-\frac{1}{\epsilon^2}=(z+1)(z-1)(z+\frac{i}{\epsilon})(z-\frac{i}{\epsilon})$
has no multiple roots. From our general results in previous sections, the ODE (\ref{ODE-example2}) 
has polynomial solutions of degree $n=1,2\cdots,$
\beq
y(z)=\prod_{i=1}^n(z-z_i),
\eeq
where $z_i$ are the roots of the above polynomial to be determined, provided that $E, \epsilon$ satisfy the relations
\beqa
&&E=\frac{\mu^2}{4}(n-1)(5-n),\label{E-gheun1}\\
&&\sum_{i=1}^nz_i=0,\\
&&\frac{6(n-1)}{\epsilon^2}=(n-1)(n-6)+[5-2(n-1)]\sum_{i=1}^nz_i^2-2\sum_{i<j}^nz_iz_j,\label{epsilon}
\eeqa
and the Bethe ansatz equations,
\beq
\left[z_i^4+\left(\frac{1}{\epsilon^2}-1\right)z_i^2-\frac{1}{\epsilon^2}\right]\sum_{j\neq i}^n\frac{2}{z_i-z_j}
    =5z_i^3-\left(\frac{1}{\epsilon^2}+6\right)z_i,~~~~i=1,2,\cdots,n.
\eeq

For $n=1$, we have $E=0$ and $z_1=0$ so that $y(z)=z$. There is no constraint on $\epsilon$, and 
\beq
\psi(x)= \frac{\cosh\frac{\mu x}{2}}{\left(1+\frac{1}{\epsilon^2}+\sinh^2\frac{\mu x}{2}\right)^{\frac{3}{2}}}
\eeq
which is the ground state of the system with energy eigenvalue $E=0$.

For $n=2$, we have $E=\frac{3}{4}\mu^2$ and equations,
\beqa
&&\frac{6}{\epsilon^2}=-4+3(z_1^2+z_2^2)-2z_1z_2,~~~~~~z_1=-z_2,\\
&& \left[z_1^4+\left(\frac{1}{\epsilon^2}-1\right)z_1^2-\frac{1}{\epsilon^2}\right]\frac{2}{z_1-z_2}
    =5z_1^3-\left(\frac{1}{\epsilon^2}+6\right)z_1,\\
&&\left[z_2^4+\left(\frac{1}{\epsilon^2}-1\right)z_2^2-\frac{1}{\epsilon^2}\right]\frac{2}{z_2-z_1}
    =5z_2^3-\left(\frac{1}{\epsilon^2}+6\right)z_2,
\eeqa
which give $z_1=-z_2=\sqrt{2}$ and $\frac{1}{\epsilon^2}=2$. Thus $y(z)=(z-\sqrt{2})(z+\sqrt{2})=z^2-2$ and
\beq
\psi(x)=\frac{\cosh^2\frac{\mu x}{2}-2}{\left(1+\frac{1}{\epsilon^2}+\sinh^2\frac{\mu x}{2}\right)^{\frac{3}{2}}}.
\eeq
This gives the first excited state eigenfunction. The two energy eigenvalues and eigenfunctions 
above for $n=1, 2$ reproduce those obtained in \cite{CL75} (from a completely different analysis). Here, we have
obtained the closed form expressions for all eigenvalues and eigenfunctions of the system. From (\ref{E-gheun1}),
we see that there are only five non-negative energy solutions to the Schr\"odinger equation with energy eigenvalues $E\leq \mu^2$.
All other analytic solutions (corresponding to $n\geq 6$) have negative energy eigenvalues. These negative energy
solutions correspond to unstable modes.

\subsection{Static perturbations for the non-extremal Reissner-Nordstr\"om solution}

Consider the Lagrangian describing 4-dimensional gravity coupled to an abelian  
gauge field $A_\mu$ and a real massive scalar $\phi$ (with mass $m_s$) \cite{Gubser05}, 
\beqa
&&{\cal L}=\sqrt{g}\left[\frac{R}{16\pi G}-\frac{1}{2}(\partial_\mu\phi)^2-\frac{f(\phi)}{4}F_{\mu\nu}^2-V(\phi)\right],\n
&&V(\phi)=\frac{1}{2}m_s^2\phi^2,~~~~~f(\phi)=\frac{1}{1+a^2\phi^2},
\eeqa
where $R$ is the scalar curvature, $G$ is the Newton constant, and $a$ is a parameter 
which has the dimension of length. The interaction between the gauge and scalar fields is 
non-renormalizable. For static solution with magnetic charge $g_m$, defined by $\int_{S^2}F=4\pi g_m$, where
$F=\frac{1}{2}F_{\mu\nu}dx^\mu \wedge dx^\nu$ is the gauge field 2-form, one must have \cite{Gubser05}
\beqa
&&ds^2=g_{tt}dt^2+g_{rr}dr^2+r^2(d\theta^2+\sin^2\theta\,d\varphi^2),\n
&&F=g_m\,d\theta \wedge \sin\theta\,d\varphi,~~~~~~\phi=\phi(r).
\eeqa
The relevant equations of motion are
\beqa
\Box \phi&=&\frac{1}{\sqrt{g}}\partial_r\sqrt{g}g^{rr}\partial_r\phi=\frac{\partial\ V_{\rm eff}}{\partial \phi},~~~~~
 V_{\rm eff}(\phi, r)=V(\phi)+\frac{g_m^2}{2r^4}f(\phi),\n
G_{\mu\nu}&=&8\pi G\, T_{\mu\nu}.
\eeqa
By expressing $g_{tt}=-e^{2A(r)}$ and $g_{rr}=e^{2B(r)}$, one obtains (in the unit $1/\sqrt{8\pi G}=1$) 
the following system of equations for $\phi$ and $B$,
\beqa
&&\phi''+\left(\frac{2}{r}-2B'+\frac{r}{2}\phi'^2\right)=e^{2B} \frac{\partial\ V_{\rm eff}}{\partial \phi},\n
&&\frac{1}{2} \phi'^2+e^{2B}V_{\rm eff}-\frac{2}{r}B'+\frac{1-e^{2B}}{r^2}=0.
\eeqa
Linearizing  the above equations around $\phi=0$ and fitting $e^{-2B}$ to the Reissner-Nordstr\"om
form for large $r$ gives rise to \cite{Gubser05}
\beqa
&&\phi''+\left(\frac{2}{r}-2B'\right)\phi'=e^{2B}\,\frac{\partial^2\ V_{\rm eff}}{\partial \phi^2}(0,r)\phi,\n
&&B(r)=-\frac{1}{2}\log\left(1-\frac{M}{4\pi r}+\frac{g_m^2}{2r^2}\right),~~~~
 \frac{\partial^2\ V_{\rm eff}}{\partial \phi^2}(0,r)=m_s^2-\frac{g_m^2a^2}{r^4},\n
&&\phi\propto 1+\frac{2(g_m^2a^2-m_s^2)}{g_m^2-2}(r-1)~~~{\rm for}~~~r\rightarrow 1,\n
&&\phi\propto e^{-m_sr}~~~{\rm for}~~~r\rightarrow\infty,
\eeqa
where $M$ is the mass of the Reissner-Nordstr\"om black hole. In the non-extremal case $M^2 > 32\pi^2g_m^2$, the function 
$r^2-\frac{M}{4\pi}r+\frac{g_m^2}{2}$ has two roots
\beq
r_\pm =\frac{M}{8\pi}\pm \frac{1}{8\pi}\sqrt{M^2-32\pi^2 g_m^2},
\eeq
which correspond to the two horizons of the black hole. Noting that $r_\pm$ obey the relations,
\beq
r_+\,r_-=\frac{g_m^2}{2},~~~~~r_+ + r_-=\frac{M}{4\pi}, \label{r+r-relations}
\eeq
then the ODE for $\phi$ can be brought into the form,
\beq
\phi''+p(r) \phi'+q(r)\phi=0.\label{linearizedODE}
\eeq
Here
\beqa
p(r)&=&\frac{1}{r-r_+}+\frac{1}{r-r_-},\n
q(r)&=&-m_s^2+\frac{2a^2}{r^2}+2a^2(\frac{1}{r_+}+\frac{1}{r_-})\frac{1}{r}
      +\frac{q_+}{r-r_+} +\frac{q_-}{r-r_-}
\eeqa
with
\beqa
q_+&=&\frac{1}{r_+-r_-}\left(  m_s^2r_+^2-\frac{4a^2r_-^2}{g_m^2}+\frac{m_s^2g_m^2}{2}(1-r_++r_-)\right),\n
q_-&=&\frac{1}{r_+-r_-}\left(   m_s^2r_-^2+\frac{4a^2r_+^2}{g_m^2}+\frac{m_s^2g_m^2}{2}(1-r_++r_-)\right).
\eeqa
The Lagrangian has a rigid rescaling symmetry. This scaling freedom can be used to set the horizon location $r_+=1$.
Then black hole mass $M$ can be determined by requiring that the horizon is at $r_+ =1$. 
It follows from (\ref{r+r-relations}) that $M=2\pi\left(g_m^2+2\right)$ and $r_-=\frac{g_m^2}{2}$. The corresponding
$q(r)$ and $p(r)$ reduce to  \footnote{Authors in \cite{BSW06} also considered the differential equation (\ref{linearizedODE}) but arrived at
the different values of $M, r_-$ as well as the different $p(r), q(r)$ functions.} 
\beqa
p(r)&=&\frac{1}{r-1}+\frac{1}{r-r_-},\n
q(r)&=&-m_s^2+\frac{2a^2}{r^2}+2a^2(1+\frac{1}{r_-})\frac{1}{r}\n
  & & +\frac{a^2g_m^2-m_s^2(1+r_-^2)}{1-r_-}\frac{1}{r-1}
      +\frac{m_s^2g_m^2r_+ + \frac{2a^2}{r_-}}{1-r_-}\frac{1}{r-r_-}.\label{prqr-at-horizon}
\eeqa
We now find exact solutions to (\ref{linearizedODE}) with $q(r),~p(r)$ given by (\ref{prqr-at-horizon}). By means of
transformation
\beq
\phi(r)=r^\mu\,e^{-m_sr}\,f(r),~~~~~\mu=\frac{1}{2}\left(1\pm \sqrt{1-8a^2}\right),
\eeq
then it can be shown that $f(r)$ satisfies the following ODE
\beq
f''+\left(\frac{2\mu}{r}+\frac{1}{r-1}+\frac{1}{r-r_-}-2m_s\right)f'+\frac{c_2r^2+c_1r+c_0}{r(r-1)(r-r_-)}f=0,\label{gheun2-appl1}
\eeq
where
\beqa
c_2&=&2a^2(1+\frac{1}{r_-})-2m_s(\mu+1)+\frac{1}{1-r_-}\left(g_m^2(a^2+m_s^2r_-)-m_s^2(1+r_-^2)+\frac{2a^2}{r_-}\right),\n
c_1&=&\left[m_s(2\mu+1)+\mu\right](r_-+1)-\frac{2a^2(r_-+1)^2}{r_-}\n
   & &-\frac{1}{1-r_-}\left(g_m^2(a^2+m_s^2)r_--m_s^2(1+r_-^2)r_-+\frac{2a^2}{r_-}\right),\n
c_0&=&2a^2+2\left[a^2-\mu(m_s+1)\right]r_-.\label{c2c1c0-for-RN}
\eeqa
This ODE is of the form (\ref{gHeun2}). Applying our general results in previous sections, we can show that (\ref{gheun2-appl1})
 has polynomial solutions of degree $n=0,1,2,\cdots,$
\beq
f(r)=\prod_{i=1}^n(r-r_i),~~~~~f\equiv 1~{\rm for}~n=0,
\eeq
where $r_i$ are the roots of the above polynomial to be determined, provided that $a, m_s, g_m$ obey the following relations:
\beqa
&&2a^2(1+\frac{1}{r_-})-2m_s(\mu+1)\n
&&~~~~~~~~~+\frac{1}{1-r_-}\left(g_m^2(a^2+m_s^2r_-)-m_s^2(1+r_-^2)+\frac{2a^2}{r_-}\right)
 =2m_sn,\label{c2-gheun2-appl}\\
&&\left[m_s(2\mu+1)+\mu\right](r_-+1)-\frac{2a^2(r_-+1)^2}{r_-}\n
&&~~~~~~~~-\frac{1}{1-r_-}\left(g_m^2(a^2+m_s^2)r_--m_s^2(1+r_-^2)r_-+\frac{2a^2}{r_-}\right)\n
&&~~~~~~~=2m_s\sum_{i=1}^nr_i-n[n+2\mu+1+2m_s(r_-+1)],\label{c1-dheun2-appl}\\
&&2a^2+2\left[a^2-\mu(m_s+1)\right]r_- \n
&&~~~~~~~= 2m_s\sum_{i=1}^nr_i^2-2\left[n+\mu+m_s(r_-+1)\right]\sum_{i=1}^nr_i\n
&&~~~~~~~~+n\left[(n-2\mu)(r_-+1)-2m_sr_-\right],\label{c0-gheun2-appl}
\eeqa
where the roots $r_i$ are determined by the set of Bethe ansatz equations,
\beq
\sum_{j\neq i}^n\frac{2}{r_i-r_j}+\frac{2\mu}{r_i}+\frac{1}{r_i-1}+\frac{1}{r_i-r_-}-2m_s=0,~~~~i=1,2,\cdots,n.
\eeq

Note that $f(r)=1$ is a solution of the ODE (\ref{gheun2-appl1}) provided that $a, m_s, g_m$ satisfy the following
relations
\beqa
&&2a^2(1+\frac{1}{r_-})-2m_s(\mu+1)+\frac{1}{1-r_-}\left(g_m^2(a^2+m_s^2r_-)-m_s^2(1+r_-^2)+\frac{2a^2}{r_-}\right)=0,\n
&&\left[m_s(2\mu+1)+\mu\right](r_-+1)-\frac{2a^2(r_-+1)^2}{r_-}\n
&&~~~~~~~~~~~~  -\frac{1}{1-r_-}\left(g_m^2(a^2+m_s^2)r_--m_s^2(1+r_-^2)r_-+\frac{2a^2}{r_-}\right)=0,\n
&&2a^2+2\left[a^2-\mu(m_s+1)\right]r_-=0. \label{f(r)=1}
\eeqa
These relations can be obtained from (\ref{c2-gheun2-appl})-(\ref{c0-gheun2-appl}) by letting $n=0$. It follows that 
\beq
\phi(r)=r^{\frac{1}{2}(1\pm\sqrt{1-8a^2})}e^{-m_sr},
\eeq
where $a$ and $m_s$ are determined by equations (\ref{f(r)=1}). This gives the first exact solution of the differential 
equation (\ref{linearizedODE}) such that the horizon is at $r_+=1$.

\subsection{Planar Dirac electron in Coulomb and magnetic fields}

Consider the (2+1)-dimensional relativistic system of a Dirac electron (with mass $m_e$) in the presence of an external 
electromagnetic field $A_\mu$. This system was also examined in \cite{CH02} via a similar Bethe ansatz approach. 
The covariant Dirac equation (in the unit $\hbar=c=1$) has the form
\beq
i\gamma^\mu(\partial_\mu +ieA_\mu)\Psi(t, {\bf r})=m_e\Psi(t, {\bf r}),
\eeq
where $m_e$ is the rest mass of the electron, $-e~(e>0)$ is its electric charge and the $(2+1)$ Dirac 
gamma matrices $\gamma^\mu$ 
satisfy the anti-commutation relations $\{\gamma^\mu,\gamma^\nu\}=2\eta^{\mu\nu}$ 
with $\eta^{\mu\nu}=\eta_{\mu\nu}={\rm diag}(1,-1,-1)$. 
In an external Coulomb and a constant homogeneous magnetic field $B$, the vector potential can be written as 
\beq
A_0=-\frac{Ze}{r},~~~~~~A_1=-\frac{By}{2},~~~~~A_2=\frac{Bx}{2}.
\eeq
Then the Hamiltonian $H({\bf r})$ of the system can be expressed as
\beq
i\partial_0\Psi(t, {\bf r})=H({\bf r})\Psi(t, {\bf r}),~~~~~
     H({\bf r})=\gamma^0\gamma^kP_k+eA_0+\gamma^0 m_e,\label{electron-Hamiltonian}
\eeq
where $P_k=-i\partial_k+eA_k,~k=1,2,$ is the operator of generalized momentum of the electron. The wave function
$\Psi(t, {\bf r})$ is assumed to have the form
\beq
\Psi(t, {\bf r})=\frac{1}{\sqrt{r}}\exp(-iEt)\,\psi_l(r,\theta),\label{Psi-function}
\eeq
where $E$ is the energy eigenvalue of Hamiltonian, and 
\beq
\psi_l(r,\theta)=\left(
\begin{array}{c}
F(r)\,e^{il\theta}\\
G(r)\,e^{i(l+1)\theta}
\end{array}
\right),\label{psi-function}
\eeq
where $l$ is an integer. Substituting (\ref{Psi-function}) and (\ref{psi-function}) into (\ref{electron-Hamiltonian}) and working in
the polar coordinates $(t,r,\theta)$ reduce the problem to a system of coupled differential equations for $F(r)$ and $G(r)$,
\beqa
\frac{dF}{dr}-\left(\frac{l+\frac{1}{2}}{r}+\frac{eBr}{2}\right)F+\left(E+m_e+\frac{Z\alpha}{r}\right)G&=&0,\\
\frac{dG}{dr}+\left(\frac{l+\frac{1}{2}}{r}+\frac{eBr}{2}\right)G+\left(E-m_e+\frac{Z\alpha}{r}\right)F&=&0,
\eeqa
where $\alpha=e^2=1/137$ is the fine structure constant. Solving $G(r)$ from the first equation in terms of $F(r)$ and substituting
into the second equation, we obtain the second-order ODE for $F(r)$,
\beqa
&&F''+\left(\frac{1}{r}-\frac{1}{r+r_0}\right)F'+\left\{E^2-m_e^2-eB(l+1)+\frac{2EZ\alpha+(l+\frac{1}{2})/r_0}{r}\right.\n
&&~~~~~~~~~~~~~\left.-\frac{\frac{eB}{2}r_0+(l+\frac{1}{2})/r_0}{r+r_0}-\frac{(l+\frac{1}{2})^2-(Z\alpha)^2}{r^2}-\frac{(eB)^2}{4}r^2\right\}F=0,
\eeqa
where $r_0=\frac{Z\alpha}{E+m_e}$. Applying the transformation,
\beq
F(r)=r^\xi\,e^{-eB\frac{r^2}{4}}\,f(r),~~~~~~\xi=\sqrt{(l+\frac{1}{2})^2-(Z\alpha)^2},\label{Ff-transformation}
\eeq
we obtain
\beq
f''+\left(\frac{2\xi+1}{r}-\frac{1}{r+r_0}-eBr\right)f'+\frac{c_2r^2+c_1r+c_0}{r(r+r_0)}f=0,\label{gheun3-appl1}
\eeq
where
\beqa
c_2&=&E^2-m_e^2-eB(\xi+l+\frac{3}{2}),\n
c_1&=&2EZ\alpha+\left[E^2-m_e^2-eB(\xi+l+\frac{5}{2})\right]r_0,\n
c_0&=&2EZ\alpha r_0+l+\frac{1}{2}-\xi.
\eeqa
This ODE is of the form (\ref{gHeun3}) and has, from our general results in previous sections, polynomial solutions of degree $n=0,1,2,\cdots,$
\beq
f(r)=\prod_{i=1}^n(r-r_i),~~~~~f\equiv 1~{\rm for}~n=0,
\eeq
where $r_i$ are the roots of the above polynomial to be determined, provided that $E, Z, B$ are given by
\beqa
&&E^2=m_e^2+eB\left(n+l+\xi+\frac{3}{2}\right),\n
&&2EZ\alpha=eB\left(r_0+\sum_{i=1}^{n}r_i\right),\n
&&2EZ\alpha r_0=-n(n+2\xi-1)+\xi-(l+\frac{1}{2})+eB\left(\sum_{i=1}^nr_i^2+r_0\sum_{i=1}^nr_i\right), 
         \label{results--gheun3}
\eeqa
and the Bethe ansatz equations
\beq
\sum_{j\neq i}^n\frac{2}{r_i-r_j}+\frac{2\xi+1}{r_i}-\frac{1}{r_i+r_0}-eBr_i=0,~~~~~i=1,2,\cdots,n.
         \label{BAEs-DiracElectron}
\eeq

Let us remark that our (\ref{results--gheun3})-(\ref{BAEs-DiracElectron}) differ from the corresponding 
equations (22)-(25) of ref.\cite{CH02}. It would be interesting to establish the relation between the two sets of expressions.

Note that $f(r)=1$ is a solution of the ODE (\ref{gheun3-appl1}) provided that $E, Z, B$ obey the following relations:
\beqa
&&E^2=m_e^2+eB\left(l+\xi+\frac{3}{2}\right),\\
&&2EZ\alpha=eBr_0,\n
&&2EZ\alpha r_0=\xi-(l+\frac{1}{2}).
\eeqa
These relations can be obtained from (\ref{results--gheun3}) by setting $n=0$. Solving these relations we obtain
\beqa
eB&=&-\frac{m_e^2(l+\frac{1}{2}+\xi)}{(l+1+\xi)^2},\n
E&=&-\frac{m_e}{2}+\frac{1}{2}\sqrt{m_e^2+2eB}=-\frac{m_e}{2(l+1+\xi)},\label{eBE-n=0}
\eeqa
where $\xi$ is related to the parameter $Z$ via the expression given in (\ref{Ff-transformation}).
It follows that
\beq
F(r)=r^\xi\, e^{-eB\frac{r^2}{4}},~~~~G(r)=\frac{\xi-l-\frac{1}{2}+eBr^2}{(E+m_e)r+Z\alpha}\,r^\xi\, e^{-eB\frac{r^2}{4}}\label{exact-FG}
\eeq
with $eB$ and $E$ given by (\ref{eBE-n=0}) in terms of $\xi$ (i.e. $Z$). As far as we know, 
(\ref{exact-FG}) gives the first exact solution to the planar Dirac electron system. 
However, this solution does not seem to be squarely  integrable.

\subsection{Schr\"odinger equation of $O(N)$ invariant decatic anharmonic oscillator}

Consider the $O(N)$ invariant decatic anharmonic oscillator in $N$ dimensions \cite{Pan99}. The Schr\"odinger equation is
\beq
\left(-\frac{1}{2}\nabla^2+V({\bf x})\right)\psi({\bf x})=E\psi({\bf x})
\eeq
with the potential $V({\bf x})$ defined by
\beq
V({\bf x})=\lambda_1{\bf x}^2+\lambda_2\left({\bf x}^2\right)^2+\lambda_3\left({\bf x}^2\right)^3
  +\lambda_4\left({\bf x}^2\right)^4+\left({\bf x}^2\right)^5,~~~~{\bf x}^2=\sum_{i=1}^Nx_i^2.
\eeq  
Here without loss of generality we have normalized the potential such that the coefficient of 
$\left({\bf x}^2\right)^5$ equals to 1.
In $N$-dimensional spherical coordinates, the radial wave function $R(r)$ satisfies
\beq
\frac{1}{2}\left[-\frac{d^2}{dr^2}-\frac{N-1}{r}\frac{d}{dr}+\frac{l(l+N-2)}{r^2}+2(V(r)-E)\right]R(r)=0.\label{diff-for-R}
\eeq
Applying the transformation $R(r)=r^{\frac{1-N}{2}} \psi(r)$ yields
\beq
-\psi''+\left[\frac{\mu(\mu-1)}{r^2}+2(V(r)-E)\right]\psi=0,
\eeq
where $\mu=l+\frac{1}{2}(N-1)$. Applying the transformation
\beq
\psi=r^\mu\,e^{-\alpha\frac{r^2}{2}-\beta\frac{r^4}{4}-\gamma\frac{r^6}{6}}\phi,~~~~~
z=r^2,
\eeq
where $\alpha, \beta, \gamma>0$ are parameters yet to be determined, yields the ODE for $\phi(z)$ 
\beqa
&&\phi''+\left(\frac{l+N/2}{z}-\gamma z^2-\beta z-\alpha \right)\phi'+\frac{1}{4z}\left[\left(2\alpha\beta-\gamma(N+2l+4)-2\lambda_2\right)z^2\right.\n
&&~~~~~~~~~~~~~\left.
    +\left(\alpha^2-\beta(N+2l+2)-2\lambda_1\right)z+2E-\alpha(N+2l)\right]\phi=0, \label{ODE-gHeun4}
\eeqa
where $\alpha, \beta, \gamma$ are given by
\beq
\gamma=\sqrt{2},~~~~\beta=\frac{\lambda_4}{\sqrt{2}},~~~~\alpha=\frac{1}{\sqrt{2}}\left(
   \lambda_3-\frac{\lambda_4^2}{4}\right)
\eeq
The ODE (\ref{ODE-gHeun4}) is of the form (\ref{gHeun4}). Applying our general results in previous section, we obtain that
this equation has polynomial solutions of degree $n=0,1,2,\cdots,$
\beq
\phi(z)=\prod_{i=1}^n(z-z_i),~~~~~\phi\equiv 1~{\rm for}~n=0,
\eeq
where $z_i$ are the roots of the above polynomial to be determined, provided that
\beqa
&&2\alpha \beta-\gamma(N+2l+4)-2\lambda_2=4\gamma n,\label{c2-gHeun4}\\
&&\alpha^2-\beta(N+2l+2)-2\lambda_1=4\gamma\sum_{i=1}^nz_i+4\beta n,\label{c1-gHeun4}\\
&&E=\frac{\alpha}{2}(4n+N+2l)+2\beta\sum_{i=1}^nz_i+2\gamma\sum_{i=1}^nz_i^2,\label{c0-gHeun4}
\eeqa
and the Bethe ansatz equations,
\beq
\sum_{j\neq i}^n\frac{2}{z_i-z_j}=\gamma z_i^2+\beta z_i+\alpha -\frac{l+N/2}{z_i},~~~~i=1,2,\cdots, n.\label{BAEs-gHeun4}
\eeq

Note that $\phi=1$ is a solution to (\ref{ODE-gHeun4}) with energy
\beq
E=\frac{1}{2\sqrt{2}}\left(\lambda_3-\frac{\lambda_4^2}{4}\right)(N+2l),
\eeq
where $\lambda_3$ and $\lambda_4$ obey the constraints,
\beqa
&&\lambda_4\left(\lambda_3-\frac{\lambda_4^2}{4}\right)-\sqrt{2}(N+2l+4)-2\lambda_2=0,\\
&&\frac{1}{2}\left(\lambda_3-\frac{\lambda_4^2}{4}\right)^2-\frac{\lambda_4}{\sqrt{2}}(N+2l+2)-2\lambda_1=0,
\eeqa
which can be obtained from (\ref{c2-gHeun4})-(\ref{c0-gHeun4}) by setting  $n=0$. The real 
solutions of these constraint equations (so that the energy $E$ is real) are given by
\beqa
\lambda_4&=&-\frac{2\sqrt{2}\lambda_1}{3(N+2l+2)}+\left(-\frac{q}{2}+\sqrt{(\frac{q}{2})^2+(\frac{p}{3})^3}\right)^{1/3}\n
 & &  +\left(-\frac{q}{2}-\sqrt{(\frac{q}{2})^2+(\frac{p}{3})^3}\right)^{1/3},\n
\lambda_3&=&\frac{\lambda_4^2}{4}+\frac{\sqrt{2}\left(N+2l+4+\sqrt{2}\lambda_2\right)}{\lambda_4},\label{lambda3lambda4-n=0}
\eeqa
where
\beqa
p&=&-\frac{24\lambda_1^2}{9(N+2l+2)^2},\n
q&=&\frac{32\sqrt{2}\lambda_1^3}{27(N+2l+2)^3}-\frac{(N+2l+4+\sqrt{2}\lambda_2)^2}{N+2l+2}.
\eeqa
We thus obtain the first exact ground state of the radial Schr\"odinger equation (\ref{diff-for-R}), 
\beq
R(r)=r^{l}\,\exp\left[-\frac{1}{2\sqrt{2}}\left(\lambda_3-\frac{\lambda_4^2}{4}\right)r^2
    -\frac{\lambda_4}{4\sqrt{2}}\,r^4-\frac{1}{3\sqrt{2}}\,r^6\right]
\eeq    
with $\lambda_3$ and $\lambda_4$ given by (\ref{lambda3lambda4-n=0}).

For $n=1$, we find that the real root to the Bethe ansatz equation (\ref{BAEs-gHeun4}) is
\beq
z_1=\frac{\lambda_4}{6}+\left(-\frac{v}{2}+\sqrt{(\frac{v}{2})^2+(\frac{u}{3})^3}\right)^{1/3}
   +\left(-\frac{v}{2}-\sqrt{(\frac{v}{2})^2+(\frac{u}{3})^3}\right)^{1/3},\label{z1-gheun4-for-n=1}
\eeq
where 
\beqa
u&=&\frac{\lambda_4^2}{24}-\frac{\lambda_3}{2},\n
v&=&\frac{5\lambda_4^3}{432}-\frac{1}{2}\lambda_3+\frac{1}{2\sqrt{2}}(N+2l).
\eeqa
The parameters $\lambda_3$ and $\lambda_4$ are determined from the equations (see (\ref{c2-gHeun4}) and (\ref{c1-gHeun4})),
\beqa
&&\lambda_4\left(\lambda_3-\frac{\lambda_4^2}{4}\right)-\sqrt{2}(N+2l+8)-2\lambda_2=0,\\
&&\frac{1}{2}\left(\lambda_3-\frac{\lambda_4^2}{4}\right)^2-\frac{\lambda_4}{\sqrt{2}}(N+2l+6)-2\lambda_1=4\sqrt{2}z_1.
\eeqa
The energy eigenvalue $E$ is given by
\beq
E=\frac{1}{2\sqrt{2}}\left(\lambda_3-\frac{\lambda_4^2}{4}\right)(N+2l+4)+\sqrt{2}\lambda_4\,z_1+2\sqrt{2}\,z_1^2,
\eeq
and the radial wave function is
\beq
R(r)=r^{l}\left(r^2-z_1\right)\exp\left[-\frac{1}{2\sqrt{2}}\left(\lambda_3-\frac{\lambda_4^2}{4}\right)r^2
    -\frac{\lambda_4}{4\sqrt{2}}\,r^4-\frac{1}{3\sqrt{2}}\,r^6\right]
\eeq
with $z_1$ given by (\ref{z1-gheun4-for-n=1}). This gives the first excited state of the system.

\section{Concluding remarks}

The main results of this paper are theorem \ref{main-results}, corollary \ref{by-products} and their applications to
the exact solutions of the physical systems (examined in section \ref{applications}).

As spelled out in theorem \ref{main-results}, we have found all polynomials $Z(z)$ such that the ODE (\ref{ODE}) 
has polynomial solutions $S(z)$ of degree $n$ with distinct
roots $z_1,z_2,\cdots, z_n$. We have also found a set of $n$ algebraic 
equations which determine the roots $z_i$ and thus the corresponding polynomial solutions $S(z)$.
If the coefficients of polynomials $X(z)$ and $Y(z)$ in (\ref{ODE}) are algebraically dependent, i.e satisfy the relations
in corollary \ref{by-products}, then ODE (\ref{ODE}) allows an $sl(2)$ algebraization.  
We have also found all the polynomials $Z(z)$ and degree $n$ polynomial solutions $S(z)$ under these conditions.

Having as many as 12 parameters, the ODE (\ref{ODE}) is very general and contains, as special cases, most known second order
differential equations  which occur in physical, chemical and engineering applications in the literature 
(e.g. the relatively simple Heun equation and its various confluent equations). 
Thus our general results (theorem \ref{main-results} and
its corollaries) provide an unified derivation of exact, closed form solutions for all such systems. 
As applications to theorem \ref{main-results}, we have examined five physical systems in section \ref{applications},
which are described by the ODEs corresponding to (\ref{Heun})-(\ref{gHeun4}), respectively. We have shown that these systems 
are quasi-exactly solvable, i.e. the corresponding differential equations have polynomial solutions, 
if their parameters satisfy certain constraints (special cases of (\ref{c2-expression})-(\ref{c0-expression})). 
The quasi-exact solvability has enabled us to use theorem \ref{main-results} to
obtain the closed form expressions for the eigenvalues and eigenfunctions of these systems.

Our results (theorem \ref{main-results} and corollary \ref{by-products}) can be extended to second order ODE of the form
(\ref{ODE}) with deg$X(z)\leq l$, deg$Y(z)\leq l-1$ and deg$Z(z)\leq l-2$  for  $l\geq 5$ as well as to higher order ODEs.
Research on this as well as on applications of theorem \ref{main-results} and corollary \ref{by-products} in 
various areas of science is in progress, and results will be reported elsewhere.

\vskip.3in
\noindent {\bf Acknowledgments:} I would like to thank Ryu Sasaki for a very careful reading of the manuscript 
and many critical comments, and Clare Dunning for a careful reading of the manuscript and helpful suggestions. 
I also thank Tony Bracken, G\"unter von Gehlen, Mark Gould, and Jon Links for 
comments and suggestions. This work was supported by the Australian Research Council.

\sect{Appendix}

For completeness and convenience of applications, in this appendix we write down the explicit formulas obtained 
from applying the theorem \ref{main-results} to the special cases (\ref{Heun})-(\ref{gHeun4}).  

\vskip.1in
\begin{Corollary} The coefficients $c_1, c_0$ of $Z(z)$ such that the Heun equation (\ref{Heun}) has polynomial 
solution (\ref{polynomial-S}) are given by
\beqa
c_1&=& -n\left[n-1+\sum_{s=1}^3\alpha_s\right],\label{c1-heun}\\
c_0&=&-\left[2(n-1)+\sum_{s=1}^3\alpha_s\right]\sum_{i=1}^nz_i+n(n-1)\sum_{s=1}^3d_s\n
& &+n[\alpha_1(d_2+d_3)+\alpha_2(d_1+d_3)+\alpha_3(d_1+d_2)],\label{c0-heun}
\eeqa
where the roots $z_1,z_2,\cdots, z_n$ are determined by the Bethe ansatz equations,
\beq
\sum_{j\neq i}^n\frac{2}{z_i-z_j}+\sum_{s=1}^3\frac{\alpha_s}{z_i-d_s}=0,~~~i=1, 2,\cdots,n.\label{BAEs-heun}
\eeq
\end{Corollary}
Write $c_1=\alpha\beta$. Then (\ref{c1-heun}) is nothing but the so-called Fuchsian relation, 
$ %\beq
\alpha+\beta+1=\sum_{s=1}^3\alpha_s,
$ %\eeq
where $\alpha=-n$ and $\beta=\sum_{s=1}^3\alpha_s+n-1$. 

\vskip.1in
\begin{Corollary} The coefficients $c_2, c_1, c_0$ of $Z(z)$ such that the generalized Heun equation (\ref{gHeun1}) 
has polynomial solution (\ref{polynomial-S}) are
\beqa
c_2&=&-n\left(\sum_{s=1}^4\mu_s+n-1\right),\label{c2-gheun1}\\
c_1&=& -\left(\sum_{s=1}^4\mu_s+2(n-1)\right)\sum_{s=1}^n z_i+n\left[(n-1)\sum_{s=1}^4e_s+P\right],\label{c1-gheun1}\\
c_0&=&-\left(\sum_{s=1}^4\mu_s+2(n-1)\right)\sum_{s=1}^n z_i^2+2\sum_{i<j}^nz_iz_j\n
& & -\left[2(n-1)\sum_{s=1}^4e_s+P \right]\sum_{i=1}^nz_i+n(n-1)\sum_{s<t}^4e_se_t+Q n,\label{c0-gheun1}
\eeqa
where 
\beqa
P&=&\mu_1(e_2+e_3+e_4)+\mu_2(e_1+e_3+e_4)+\mu_3(e_1+e_2+e_4)+\mu_4(e_1+e_2+e_3),\n
Q&=&\mu_1(e_2e_3+e_2e_4+e_3e_4)+ \mu_2(e_1e_3+e_1e_4+e_3e_4)\n
  & &+\mu_3(e_1e_2+e_1e_4+e_2e_4)+\mu_4(e_1e_2+e_1e_3+e_2e_3),\nonumber
\eeqa
and the roots $z_1, z_2,\cdots,z_n$ are determined by the Bethe ansatz equations,
\beq
\sum_{j\neq i}^n\frac{2}{z_i-z_j}+\sum_{s=1}^4\frac{\mu_s}{z_i-e_s}=0,~~~i=1, 2,\cdots,n.\label{BAEs-gheun1}
\eeq
\end{Corollary}
%Write $c_2=\alpha\beta$. Then (\ref{c2-gheun1}) is nothing but the Fuchsian relation, 
%$ %\beq
%\alpha+\beta+1=\sum_{s=1}^4\mu_s,
%$ %\eeq
%where $\alpha=-n$ and $\beta=\sum_{s=1}^4\mu_s+n-1$. 

\vskip.1in
\begin{Corollary} The coefficients $c_2, c_1, c_0$ of $Z(z)$  such that the differential equation (\ref{gHeun2}) 
has polynomials solution (\ref{polynomial-S}) are given by
\beqa
c_2&=&-n\nu,\label{c2-gheun2}\\
c_1&=& -\nu\sum_{s=1}^n z_i-n\left[(n-1)+\sum_{s=1}^3\nu_s-\nu\sum_{s=1}^3f_s\right],\label{c1-gheun2}\\
c_0&=&-\nu\sum_{s=1}^n z_i^2-\left[2(n-1)+\sum_{s=1}^3\nu_s-\nu\sum_{s=1}^3f_s\right]\sum_{i=1}^nz_i
      +n(n-1)\sum_{s=1}^3f_s\n
& &+ \left[\nu(f_1f_2+f_1f_3+f_2f_3)-\nu_1(f_2+f_3)-\nu_2(f_1+f_3)-\nu_3(f_1+f_2)\right]n,\label{c0-gheun2}
\eeqa
where the roots $z_1, z_2,\cdots, z_n$ satisfy the Bethe ansatz equations,
\beq
\sum_{j\neq i}^n\frac{2}{z_i-z_j}+\sum_{s=1}^3\frac{\nu_s}{z_i-f_s}+\nu=0,~~~i=1, 2,\cdots,n.\label{BAEs-gheun2}
\eeq
\end{Corollary}
%Write $c_2=\alpha\beta$. Then (\ref{c2-gheun2}) is nothing but the Fuchsian relation,
%\beq
%\alpha+\beta+1=\nu,
%\eeq
%where $\alpha=\frac{\nu-1}{2}+\frac{1}{2}\sqrt{(\nu-1)^2+4\nu n}$ and 
%$\beta=\frac{\nu-1}{2}-\frac{1}{2}\sqrt{(\nu-1)^2+4\nu n}$. 

\vskip.1in
\begin{Corollary} The coefficients $c_2, c_1, c_0$ of $Z(z)$ such that the differential equation (\ref{gHeun3}) 
has polynomials solution (\ref{polynomial-S}) are
\beqa
c_2&=&-n\sigma,\label{c2-gheun3}\\
c_1&=& -\sigma\sum_{s=1}^n z_i-n\left[\kappa-\sigma(g_1+g_2)\right],\label{c1-gheun3}\\
c_0&=&-\sigma\sum_{s=1}^n z_i^2-\left[\kappa-\sigma(g_1+g_2)\right]\sum_{i=1}^nz_i\n
& &-n(n-1)-n\left[\sigma_1+\sigma_2+\sigma g_1g_2-\kappa(g_1+g_2)\right],\label{c0-gheun3}
\eeqa
where the roots $z_1, z_2,\cdots, z_n$ obey the Bethe ansatz equations,
\beq
\sum_{j\neq i}^n\frac{2}{z_i-z_j}+\frac{\sigma_1}{z_i-g_1}
 +\frac{\sigma_2}{z_i-g_2}+\sigma z_i+\kappa=0,~~~i=1, 2,\cdots,n.\label{BAEs-gheun3}
\eeq
\end{Corollary}

\vskip.1in
\begin{Corollary} The coefficients $c_2, c_1, c_0$ of $Z(z)$ such that the differential equation (\ref{gHeun4}) 
has polynomials solution (\ref{polynomial-S}) are given by
\beqa
c_2&=&-n\lambda,\label{c2-gheun4}\\
c_1&=& -\lambda\sum_{s=1}^n z_i-n(\gamma-\lambda h),\label{c1-gheun4}\\
c_0&=&-\lambda\sum_{s=1}^n z_i^2-(\gamma-\lambda h)\sum_{i=1}^nz_i
   -n(\delta-\gamma h),\label{c0-gheun4}
\eeqa
where the roots $z_1, z_2, \cdots, z_n$ are determined by the Bethe ansatz equations,
\beq
\sum_{j\neq i}^n\frac{2}{z_i-z_j}+\frac{\eta}{z_i-h}
 +\lambda z_i^2+\gamma z_i+\delta=0,~~~i=1, 2,\cdots,n.\label{BAEs-gheun4}
\eeq
\end{Corollary}

\bebb{99}

\bbit{SS80}
R. Sch\"afke and D. Schmidt, The connection problem for general linear ordinary differential equations
at two regular singular points with applications in the theory of special functions,
{\it SIAM J. Math. Anal.} {\bf 11} (1980), 848-862.

\bbit{GKM09-10}
D. G\'omez-Ullate, N. Kamran and R. Milson, An extended class of orthogonal polynomials defined by a Sturm-Liouville problem,
{\it  J. Math. Anal. Appl.} {\bf 359} (2009), 352-367;
An extension of Bochner's problem: exceptional invariant subspaces, {\it J. Approx. Theory} {\bf 162} (2010), 987-1006.

%\bbit{Quesne08}
%C. Quesne, Exceptional orthogonal polynomials, exactly solvable potentials and supersymmetry,
%{\it  J. Phys. A: Math. Theor.} {\bf 41} (2008), 392001 (6 pages).

\bbit{OS09-11}
%S. Odake and R. Sasaki,  Phys. Lett. B {\bf 679}, 414 (2009); Phys. Lett. B {\bf 684}, 173 (2010).\\
C-L. Ho, S. Odake and R. Sasaki, Properties of the exceptional ($X_\ell$) Laguerre and Jacobi 
 polynomials,   arXiv:0912.5477v3 [math-ph], and references therein.

\bbit{BS07}
J. Borcea and B. Shapiro, Root asymptotics of spectral polynomials for the Lam\'e operator,
arXiv:math/0701883v2 [math.CA].

\bbit{Shapiro-et-al-08-09}
B. Shapiro and M. Tater, On spectral polynomials of the Heun equations. I., arXiv:0812.2321v1 [math-ph].\\
B. Shapiro, K. Takemura  and M. Tater, On spectral polynomials of the Heun equations. II.,arXiv:0904.0650v1 [math-ph].

\bbit{SV03}
I. Scherbak and A. Varchenko, Critical points of functions, $sl_2$ representations, and Fuchsian differential equations
with only univalued solutions, arXiv:math/0112269v4 [math.QA].

\bbit{MTV06}
E. Mukhin, V. Tarasov and A. Varchenko, Higher Lam\'e equations and critical points of master functions,
arXiv:math/0601703v2 [math.CA].

\bbit{Turbiner96}
A. Turbiner, %Comm. Math. Phys. {\bf 118}, 467 (1988).
%Quasi-exactly-solvable differential equations, 1994, hep-th/9409068,
CRC Handbook of Lie Group Analysis of Differential Equations, Vol. {\bf 3}, Chap. 12,  
%New Trends in Theoretical Developments and Computational Methods, 
ed. N. H. Ibragimov, CRC Press, Boca Raton, FL, 1996.

\bbit{Gonzarez93}
A. Gonz\'arez-L\'opez, N. Kamran and P. Olver, Normalizability of one-dimensional quasi-exactly solvable
Schr\"odinger operators, {\it Commun. Math. Phys.} {\bf 153} (1993), 117-146.

\bbit{BD96}
C.M. Bender and G.V. Dunne, Quasi-exactly solvable systems and orthogonal polynomials,
{\it J. Math. Phys.} {\bf 37} (1996), 6-11.

\bbit{KUW96}
A. Krajewska, A. Ushveridze and Z. Walczak, Bender-Dunne orthogonal polynomials and quasi-exact solvability,
arXiv:hep-th/9601088v1.

%\bbit{Dorey99}
%P. Dorey and R. Tateo, J. Phys. A: Math. Gen. {\bf 32}, L419 (1999). 

\bbit{Ushveridze94}
A.G. Ushveridze, Quasi-exactly solvable models in quantum mechanics, Institute of Physics Publishing, Bristol, 1994.

\bbit{Heine1878}
E. Heine, Handbuch der Kugelfunctionen, Vol. {\bf 1}, pp. 472-479, D. Reimer Verlag, Berlin, 1878.

\bbit{Stieltjes1885}
T. Stieltjes, Sur certains polyn\^omes qui v\'erifient une equation diff\'erentielle lin\'eaire du second
ordre et sur la theorie des fonctions de Lam\'e, {\it Acta Math.} {\bf 8} (1885), 321-326.

\bbit{Szego39}
G. Szego, Orthogonal polynomials, American Mathematical Society, 1939.

\bbit{Wiegmann94}
P.B. Wiegmann and A.V. Zabrodin, Bethe ansatz for the Bloch electron in magnetic field, {\it Phys. Rev. Lett.} 
{\bf 72} (1994), 1890-1893; 
Algebraization of difference eigenvalue equations related to $U_q(sl_2)$,
{\it Nucl. Phys.  B} {\bf 451} (1995), 699-724.

\bbit{Sasaki09}
R. Sasaki, W.-L. Yang and Y.-Z. Zhang, Bethe ansatz solutions to quasi-exactly solvable difference equations, 
{\it SIGMA} {\bf 5} (2009), 104 (16 pages).

\bbit{Yuan10}
Y.-H. Lee, W.-L. Yang and Y.-Z. Zhang, Polynomial algebras and exact solutions of general quantum non-linear optical models I: 
Two-mode boson systems, {\it J. Phys. A: Math. Theor.} {\bf 43} (2010), 185204 (17 pages);
Polynomial algebras and exact solutions of general quantum non-linear optical models II: Multi-mode boson systems,
{\it J. Phys. A: Math. Theor.} {\bf 43} (2010), 375211 (12 pages). 

\bbit{Yuan11}
Y.-H. Lee, J.R. Links and Y.-Z. Zhang, Exact solutions for a family of spin-boson systems, {\it Nonlinearity} 
{\bf 24} (2011), 1975-1786. 

\bbit{LG09}
P-F. Loos and P.M.W. Gill, Two electrons on hypersphere: a quasi-exactly solvable model, {\it Phys. Rev. Lett.}
{\bf 103} (2009), 123008 (4 pages); 
Excited states of spherium, {\it Mol. Phys.} {\bf 108} (2010), 2527-2532.

\bbit{CL75}
N.H. Christ and T.D. Lee, Quantum expansion of soliton solutions, {\it Phys. Rev. D} {\bf 12} (1975), 1606-1627.

\bbit{JKK89}
D.P. Jatkar, C.N. Kumar and A. Khare, A quasi-exactly solvable problem without $sl(2)$ symmetry,
{\it Phys. Lett. A} {\bf 142} (1989), 200-202, and references therein.

\bbit{Gubser05}
S.S. Gubser, Phase transitions near black hole horizons, {\it Class. Quant. Grav.} {\bf 22} (2005), 5121-5143.

\bbit{BSW06}
D. Batic, H. Schmid and M. Winklmeier, The generalized Heun equation in QFT in curved space-time,
{\it J. Phys. A: Math. Gen.} {\bf 39} (2006), 12559-12564.

\bbit{CH02}
C.-M. Chiang and C.-L. Ho, Planar Dirac electron in Coulomb and magnetic fields, {\it J. Math. Phys.}
{\bf 43} (2002), 43-51.

\bbit{Pan99}
F. Pan, J.R. Klauder and J.P. Draayer, Quasi-exactly solvable case of an$N$-dimensional symmetric
decatic anharmonic oscillator, {\it Phys. Lett. A} {\bf 262} (1999), 131-136. %\\
%J.R. Klauder, J. Math. Phys. {\bf 6}, 1666 (1965).

\eebb

\end{document}